\documentclass[10pt,aps,prd,showpacs,nofootinbib,groupedaddress]{revtex4-1}
\usepackage{hyperref}
\usepackage{amsfonts}
\usepackage{amsmath}
\usepackage{amssymb}
\usepackage{graphicx}
\usepackage{color}
\usepackage{textcomp}

\def\be{\begin{equation}}
\def\te{\end{equation}}
\def\ee{\end{equation}}
\def\ba{\begin{eqnarray}}
\def\bea{\begin{eqnarray}}
\def\nn{\nonumber\\}
\def\tea{\end{eqnarray}}
\def\ea{\end{eqnarray}}
\def\eea{\end{eqnarray}}

\usepackage{amssymb}

\begin{document}

\title{Not quite free shortcuts to adiabaticity}

\author{Esteban Calzetta}
\email{calzetta@df.uba.ar}
\affiliation{Departamento de F\'isica, Facultad de Ciencias Exactas y Naturales, Universidad de Buenos Aires and IFIBA, 
CONICET, Cuidad Universitaria, Buenos Aires 1428, Argentina}


\begin{abstract}
Given the increasing use of shotcuts to adiabaticity (STA) to optimize power and efficiency of quantum heat engines, it becomes a relevant question if there are any theoretical limits to their application. We argue that quantum fluctuations in the control device which implements the shortcut deflect the system from the adiabatic path. This not only induces transitions to unwanted final states but also changes the system energy, so that using the  STA has a definite cost in terms of conventional work definitions. This may be the ultimate cost of an adiabatic shortcut, in the sense that it is present even for a frictionless, zero temperature driving. We estimate the effect, to lowest nontrivial order in the derivatives of the time-dependent frequency, on a parametric harmonic oscillator, thus providing a consistency condition for the validity of the classical approximation.
\end{abstract}

\maketitle
\section{Introduction}
The option of taking \emph{shortcuts to adiabaticity} (STA's)\cite{Berry09,Torrontegui13,Campo18book} may be construed as meaning that any unitary transformation of a quantum system that may be realized adiabatically may be done arbitrarily fast too (there are also shortcuts to adiabaticity in classical mechanics \cite{Jarzynski13, Deng13}, but we shall not discuss them, see the final section). One has a quantum system in some state $\left|\psi\right\rangle $ and wants it to evolve following a preprogrammed trajectory $\left|\psi\right\rangle \left( t\right)  $. Usually this will not be a solution of the Schr\"odinger equation given the system's Hamiltonian $H_S$. We shall assume, however, that it is an approximate solution in the adiabatic limit (here ``adiabatic'' means infinitely slow; this does not imply there is no heat exchange).  If norm is preserved, then there will be some hermitian operator $H_{AS}$ (generally there will be many of them) such that 

\be 
i\hbar\frac d{dt}\left|\psi\right\rangle \left( t\right) =H_{AS}\left|\psi\right\rangle \left( t\right) 
\te 
Taking the shortcut to adiabaticity means replacing $H_S$ by $H_{AS}$; observe that the restriction to the adiabatic limit dissapears. Moreover, it usually may be arranged that not only the initial and final states, but also the Hamiltonians $H_{AS}=H_S$ at the beginning and the end of the trajectory, and so the probability distribution for the energy will be the same, whether the transformation takes place in finite time through the shortcut or in infinite time through the original Hamiltonian. This is the usual measure of work done on the system \cite{Campisi11,Augusto14,Federico17}. In this sense, it would appear that shortcuts to adiabaticity are ``free''. This would allow quantum heat machines \cite{Abah12,Campo14,Beau16,Kosloff17} to approach the ideal situation analyzed by Curzon and Ahlborn \cite{Curzon75,Zhang17}, where the only limitation to the power of the machine comes from the finite speed of heat transfer.

It is generally accepted that ``to replace $H_S$ by $H_{AS}$'' means that the system is being brought into interaction with a driving system (henceforth, ``the driving''), and that a meaningful discussion of the cost of shortcuts to adiabaticity requires including explicitly the driving within the model \cite{Zheng16}. It may be argued that the driving must have some kind of dissipation, not least to stabilize it against the backreaction from the system, and therefore that there is a cost incurred because of the need to override this dissipation. That this actually happens has been demonstrated in specific models \cite{Torrontegui17}. It has been argued that shortcuts to adiabaticity enhance work fluctuations \cite{Funo17,Abah18} along the trajectory. It is also known that there are excitations during the protocol, so that actually implementing the driving may be quite demanding \cite{Muga10}. These implementation costs are measured by time integrals of the average value of powers of 
the ``counterdiabatic'' Hamiltonian $H_1=H_{AS}-H_S$  \cite{Zheng16,Campbell17}.

Even including these likely costs, the situation of being able to drive a quantum system at arbitrary speeds with no secondary effects on the system itself is quite extraordinary. In a situation with some points in common with our subject, recently there have been proposals in the literature claiming that it was possible to cool a quantum system at a rate $\left|dT/dt\right|$ not bounded by $T$ \cite{Kolar12}, thereby in conflict with the Third Law of Thermodynamics \cite{Masanes17,Wilming17}. Closer examination showed that there was a heating effect associated with the time dependence of the fields used to drive the system, and thereby that there was an absolute lower bound to the temperature that can be reached within that class of protocols \cite{Freitas18}.

A maybe closer analogy may be drawn to the time-dependent electromagnetic fields which are used to trap cold atoms and ions. For most cold atom experiments the trapping fields can be treated with sufficient accuracy as just an external potential. However those fields fluctuate \cite{rmp}, and this causes heating of the atomic cloud over long enough time scales \cite{Savard,Gardiner}. This is one effect among several that limit the time the atoms may be kept within the trap \cite{Peng}.

We claim that, similarly, when the quantum nature of the driving is taken into account, the system deviates from the adiabatic trajectory. Moreover, the unwanted transitions' rates become  higher for faster protocols. This induces a definite deviation from the desired target in the mean energy of the system. Therefore shotcuts to adiabaticity would be ``free'' only within the approximation of treating the driving as a classical system. We point this out as a matter of principle, since the classical approximation is usually accurate \cite{Chen10,Diao18,Deng18}, but that could be relevant to a better understanding of the working and ultimate limits of shortcuts to adiabaticity. We also provide an estimate of the energy change in the system computed under the semiclassical approximation, thus providing a consistency test for the classical approximation.

For concreteness, we shall discuss shortcuts to adiabaticity in a context that is relevant to the discussion of efficiency and power of a quantum engine built from a trapped ion  \cite{Beau16,Kosloff17}. The system is modelled as a parametric oscillator \cite{Husimi53} and the desired trajectory consists on allowing the oscillator's frequency $\omega^2\left(t\right)$ to change in time without causing transitions between the instantaneous energy levels. 

If the system follows the evolution generated by its natural Hamiltonian, however, a time-dependent frequency generates excitations through parametric amplification or ``particle creation'' (\cite{Parker68,Book}). This is avoided if the changes in the frequency are infinitely slow, because in this limit a positive frequency solution remains positive frequency throughout. Moreover, in this limit the solutions are given by the so-called WKB wavefunctions. Now, the WKB approximated wave functions for the oscillator with frequency $\omega^2\left(t\right)$ are actually \emph{exact} solutions for an oscillator whose frequency changes according to a different protocol, say $\Omega^2\left(t\right)$. The necessary form for $\Omega^2\left(t\right)$ is easily computed from the original $\omega^2\left(t\right)$ \cite{Chen10b}. Thus, given any protocol $\omega^2\left(t\right)$ we can find a different protocol $\Omega^2\left(t\right)$ that would make the system follow the adiabatic trajectory of the original one.

Actually implementing the STA means that we couple our oscillator, with canonical variables $\left(x,p\right)$, to a driving, which is also a system with canonical variables $\left(\theta,\Xi\right)$, through an interaction $\Omega^2\left[\theta\right]x^2/2$, in such a way that, when $\theta$ evolves in time through the classical equations of motion for the driving, then $\Omega^2\left[\theta\left(t\right)\right]$ traces the desired protocol.

In the real world the driving will be a quantum system and there will be quantum fluctuations around the classical $\Xi$ and $\theta$. We want to know how these fluctuations in the driving affect the dynamics of the system. With this goal in mind we shall follow the evolution of the reduced Wigner function for the system to second order in the derivatives of $\Omega^2\left(\theta\right)$. If the system is initially in the $n$-th excited energy eigenstate, then, to this order, we will show that there is a finite rate for transitions to the $n\pm 2$ states, and that the final mean energy is no longer that of the $n$-th excited state of the final Hamiltonian. 

Since the transition rates are exponentially suppressed when the driving is slow, there is a regime where the classical approximation holds, as confirmed by actual experiments \cite{Muga18,CKHu18}. We regard our analysis as providing a consistency criterion for the classical approximation. In other words,  shortcuts to adiabaticity are ``free'', as measured by the difference between the system's final mean energy and the desired target, only within the classical approximation for the driving, and there are definite, if ample, limits for the validity of this approximation.

This paper is organized as follows. We present the model for system and driving in next section. Coupling to the driving turns the system into a quantum open one, and its state must be recovered as a partial trace of the system plus driving composite; in Section III we apply Feynam-Vernon Influence functional techniques to obtain the desired reduced density matrix, and in Section IV we turn this density matrix into a Wigner function through a partial Fourier transform. If the system is initialized in the $n$-th excited state, then at the end of the protocol it has a finite probability of being in the $n\pm 2$ states; this is also computed in Section IV. In Section V we estimate the actual size of the effect. We conclude with some brief final remarks. There are four appendices filling in some technical details.

\section{The model} 
As said, our system consists of a parametric oscillator. The original system Hamiltonian is 

\be
H_S=\frac{p^2}{2m}+\frac12m\omega^2\left(t\right)x^2
\te
The canonical operators $x$ and $p$  may be written as linear combinations of the initial destruction and creation operators 

\bea
x\left(t\right)&=&\sqrt{\frac{\hbar}{m}}\left\{f\left(t\right)a\left[0\right]+f^*\left(t\right)a^{\dagger}\left[0\right]\right\}\nn
p\left(t\right)&=&\sqrt{\hbar m}\left\{\dot f\left(t\right)a\left[0\right]+\dot f^*\left(t\right)a^{\dagger}\left[0\right]\right\}
\tea 
where the function $f$ solves the equation of motion 

\be 
\ddot f+\omega^2\left( t\right) f=0
\te 
with Cauchy data

\bea
f\left( 0\right) &=&\frac1{\sqrt{2\omega\left( 0\right) }}\nn
\dot f\left( 0\right) &=&-i\sqrt{\frac{\omega\left( 0\right)}2}
\tea 
Since the Wronskian $W=i\left[ f^*\dot f-f\dot f^*\right] =1$,  $f$ and $f^*$ are linearly independent. We say they form a ``particle model'', $f$ being the ``positive frequency'' solution, and $f^*$ the ``negative frequency'' one \cite{Parker68,Book}.

We assume $\dot\omega\left( 0\right) =0$. At the end of the protocol $\dot\omega\left(T\right) =0$ again, and we wish $a$ and $a^{\dagger}$ to still diagonalize the Hamiltonian $H_S$, so that a particle eigenstate at $t=0$ will still be a particle eigenstate at $T$ with the same number of particles. 

In the adiabatic limit this is the case because $f$ is given by the WKB approximation \cite{Book}

\be
f=\frac1{\sqrt{2\omega}}e^{-i\int^t\omega\left(t'\right)dt'}
\label{WKB}
\te 
One way to implement an adiabatic shortcut is to change the Hamiltonian so that the WKB wave function becomes exact. This is achieved by the Hamiltonian

\be
H_{AS}=\frac1{2m}p^2+\frac12m\Omega^2\left(t\right)x^2
\te
where 

\be
\Omega^2=\omega^2+\frac12\left(\frac{\ddot\omega}{\omega}-\frac32\left(\frac{\dot\omega}{\omega}\right)^2\right)
\label{WKB2}
\te 
We now have to write a dynamics for a composite system, made of system and driving, such that the Hamiltonian for the system will be $H_{AS}$ when the driving is treated classically. 

We model the driving as an integrable system with Hamiltonian $H_D\left[ \Xi\right] $, where $\Xi$ is an action variable; let $\theta$ be the conjugated angle variable. In the classical approximation, and neglecting back reaction, $\theta=H'_D\left[ \Xi\right] t$ evolves linearly in time. We ignore complications arising from the periodic nature of $\theta$. The full Hamiltonian is then

\be 
H=H_D+H_{AS}=H_D\left[ \Xi\right] +\frac1{2m}p^2+\frac12m\Omega^2\left(\theta\right)x^2
\label{universe}
\te
We also assume $H''_D\left[ \Xi\right] \not =0$, so we may go through the protocol at different speeds by choosing different values of $\Xi$. When $\theta$ evolves form $0$ to $2\pi$, say, $\Omega^2$ traces the desired evolution, which it completes in time $T=2\pi/H'_D\left[ \Xi\right]$

\section{STA's as quantum open systems}
Coupling the system to the driving makes the former into a quantum open system. The unitary evolution of system and driving, generated by the Hamiltonian $H$ of eq. (\ref{universe}), will entangle them. The quantum state of the system alone is the partial trace of the full density matrix with respect to the driving. We want to follow its evolution.

The proper tool for this analysis is the Feynman-Vernon influence functional \cite{FV,FH,Aurell18,Book}. We assume that at $t=0$ system and driving are uncorrelated, and the state is a direct product

\be 
\rho\left( \left( x,\theta\right)  ,\left( x',\theta'\right), 0 \right)  =\rho_{Si}\left( x,x'\right) \rho_{Di}\left( \theta,\theta'\right) 
\te 
The state at the end of the protocol will be given by a two-time path integral \cite{Book,Kamenev2011}

\bea 
&&\rho\left( \left( x,\theta\right)  ,\left( x',\theta'\right), T \right) =\int Dx^1Dx^2D\theta^1D\theta^2D\Xi^1D\Xi^2\nn
&& e^{i\left[ S\left( x^1,\theta^1,\Xi^1\right) -S^*\left( x^2,\theta^2,\Xi^2\right)\right]/\hbar }\rho_{Si}\left( x^1\left( 0\right) ,x^2\left( 0\right)\right) \rho_{Di}\left( \theta^1\left( 0\right),\theta^2\left( 0\right)\right) 
\label{ttpt}
\tea 
The trajectories in the forward branch go from $\left( x^1\left( 0\right) ,\theta^1\left( 0\right)\right)  $ to $\left( x^1\left( T\right)=x ,\theta^1\left( T\right)=\theta\right)  $; similarly for the backward branch. $\Xi^1$ and $\Xi^2$ are unconstrained. The action

\be 
S=\int\;dt\left\{\Xi\dot\theta-H_D\left[\Xi\right]+\frac12m\dot x^2-\frac12m\Omega^2\left[\theta\right]x^2\right\}
\label{action1}
\te 
has been modified in each branch to enforce path ordering.

Since our goal is to check the consistency of the classical approximation for the driving, we may assume we are in a situation where the classical approximation is expected to hold. Therefore the path integral will be dominated by trajectories that stay close to the classical evolution, $\Xi=\bar \Xi+\xi$, $\theta =\Theta+\vartheta$, with $\bar \Xi=$ constant and $\Theta=H'_D\left[ \bar \Xi\right] t$. So we expand

\be 
S=\bar S_D+\int\;dt\left\{\xi\dot\vartheta-\frac12H''_D\left[\bar\Xi\right]\xi^2+\frac12m\dot x^2-\frac12m\Omega^2\left[\Theta\right]x^2+\frac12m\left(\Omega^2\right)'\left[\Theta\right]\vartheta x^2\right\}
\te
Note that $\left(\Omega^2\right)'$ is a derivative with respect to $\Theta$. The term $\bar S_D=\int\left(\Xi\dot\Theta -H_D\left[\bar\Xi\right]\right)$ cancels out from eq. (\ref{ttpt}) because we assume the same classical trajectory in both branches. Integrating over $\xi$ we obtain

\be
S=S_S+S_D+S_{DS}
\te
where

\bea 
S_S\left( x\right) &=&\frac12\int\;dt\left\lbrace m\dot x^2+m\Omega^2\left[\Theta\left(t\right)\right] x^2\right\rbrace \nn
S_D\left( \vartheta\right) &=&\frac M2\int\;dt\;\dot \vartheta^2\nn
S_{DS}\left( x,\vartheta\right) &=&\frac m2\int\;dt\;\left( \Omega^2\right) '\left[ \Theta\right] \vartheta x^2
\label{actions}
\tea
where $M=1/H''_D\left[ \bar \Xi\right] $ is assumed to be finite. 

We obtain the state for the system by Landau tracing \cite{Landau27} over the quantum fluctuations of the driving 

\be 
\rho_{Sf}\left( x,x'\right)=\int\;d\theta\;\rho\left( \left( x,\theta\right)  ,\left( x',\theta\right), T \right)
\te 
so 

\be 
\rho_{Sf}\left( x,x'\right) =\int Dx^1Dx^2\;e^{i\left[ S_S\left( x^1\right) -S_S^*\left( x^2\right)+S_{IF}\left( x^1,x^2\right]\right] \hbar }\rho_{Si}\left( x^1\left( 0\right) ,x^2\left( 0\right)\right) 
\te 
where $S_{IF}$ is the influence action 

\be 
e^{iS_{IF}\left( x^1,x^2\right]/\hbar}=\int d\vartheta\int D\vartheta^1D\vartheta^2\;e^{i\left[S_D\left( \vartheta^1\right) + S_{DS}\left( x^1,\vartheta^1\right) -S^*_D\left( \vartheta^2\right) - S^*_{DS}\left( x^2,\vartheta^2\right)\right]/\hbar } \rho_{Di}\left( \vartheta^1\left( 0\right),\vartheta^2\left( 0\right)\right) 
\te 
The integral being over a closed time-path such that ${\vartheta^1\left( T\right) =\vartheta^2\left( T\right) =\vartheta}$ \cite{Book,Kamenev2011}. 

Our problem is to find the influence action and thereby the state of the system at $T$. The Gaussian integral over $\vartheta$ is immediate and we get 

\bea 
&&S_{IF}\left[ x^1,x^2\right] =\frac{im^2}{8\hbar}\int\;dtdt'\left( \Omega^2\right) '\left[ \Theta\left( t\right) \right] \left( \Omega^2\right) '\left[ \Theta\left( t'\right) \right] \nn
&&\left\lbrace \left( x^1\right) ^2\left( t\right) \left( x^1\right) ^2\left( t'\right)\left\langle T\left( \vartheta\left( t\right) \vartheta\left( t'\right) \right) \right\rangle +\left( x^2\right) ^2\left( t\right) \left( x^2\right) ^2\left( t'\right)\left\langle\tilde T\left( \vartheta\left( t\right) \vartheta\left( t'\right) \right) \right\rangle\right. \nn
 &-&\left. \left( x^2\right) ^2\left( t\right) \left( x^1\right) ^2\left( t'\right)\left\langle  \vartheta\left( t\right) \vartheta\left( t'\right) \right\rangle -\left( x^1\right) ^2\left( t\right) \left( x^2\right) ^2\left( t'\right)\left\langle  \vartheta\left( t'\right) \vartheta\left( t\right) \right\rangle\right\rbrace 
\tea 
where $T$ means temporal and $\tilde T$ antitemporal ordering. Writing $x^{1,2}=X\pm u/2$ this becomes

\be
S_{IF}=S_{nl}+S_d+iS_n
\te
where

\bea
S_{nl}&=&\frac{m^2}4\int\;dt\;\left(\Omega^2\left[\Theta\left(t\right)\right]\right)'u\left(t\right)X\left(t\right)
\int\;dt'\;D\left(t,t'\right)\left(\Omega^2\left[\Theta\left(t'\right)\right]\right)'X^2\left(t'\right)\nn
S_{d}&=&\frac{m^2}{16}\int\;dt\;\left(\Omega^2\left[\Theta\left(t\right)\right]\right)'u\left(t\right)X\left(t\right)
\int\;dt'\;D\left(t,t'\right)\left(\Omega^2\left[\Theta\left(t'\right)\right]\right)'u^2\left(t'\right)\nn
S_{n}&=&\frac {m^2}{2\hbar}\int\;dt\;\left(\Omega^2\left[\Theta\left(t\right)\right]\right)'u\left(t\right)X\left(t\right)
\int\;dt'\;N\left(t,t'\right)\left(\Omega^2\left[\Theta\left(t'\right)\right]\right)'X\left(t'\right)u\left(t'\right)
\tea
are associated to an induced nonlinear interaction, dissipation and noise, respectively. $D$ and $N$ are the so-called \emph{dissipation} and \emph{noise} kernels \cite{Book}

\bea
D\left(t,t'\right)&=&i\left\langle \left[\vartheta\left(t\right),\vartheta\left(t'\right)\right]\right\rangle \theta\left(t-t'\right)\nn
N\left(t,t'\right)&=&\frac12\left\langle \left\{\vartheta\left(t\right),\vartheta\left(t'\right)\right\}\right\rangle 
\tea
To compute them, write the Heisenberg $\vartheta$ operator as

\be
\vartheta =\vartheta\left(0\right)+M^{-1}P_D\left( 0\right) t
\te
so

\bea
N\left(t,t'\right)&=&\left\langle \vartheta\left(0\right)^2\right\rangle+M^{-2}\left\langle P_D\left( 0\right)^2\right\rangle tt'+\frac12M^{-1}\left\langle \left\lbrace \vartheta\left(0\right),P_D\left( 0\right)\right\rbrace \right\rangle\left(t+t'\right)\nn
D\left(t,t'\right)&=&\frac 1{2M}\left(t-t'\right)\theta \left(t-t'\right)
\label{noise}
\tea
For simplicity we shall assume that $\left\langle \left\lbrace \vartheta\left(0\right),P_D\left( 0\right)\right\rbrace \right\rangle=0$; this obtains, for example, when $\vartheta$ is initially in a Gaussian pure state (recall that by definition $\left\langle\vartheta\left(0\right)\right\rangle$ must vanish) .
We now write

\be
\mathcal{N}\left(t,t'\right)=m^2\left(\left(\Omega^2\right)'X\right) \left( t\right) N\left( t,t'\right) \left(\left(\Omega^2\right)' X\right) \left( t'\right) 
\te
and 
\be 
e^{-S_n/\hbar}=\int D\zeta\;\mathcal P\left[ \zeta, X\right] e^{i\int\;dt\;\zeta\left( t\right)u\left( t\right)/\hbar }
\te
where $\mathcal P\left[ \zeta, X\right]$ is a Gaussian measure such that $\left\langle \zeta\left( t\right) \right\rangle =0$ and

\be
\left\langle \zeta\left( t\right) \zeta\left( t'\right)\right\rangle =\mathcal{N}\left( t,t'\right)
\te
Then 

\be
e^{\left(iS_d-S_n\right)\hbar}=\int D\zeta\;\mathcal P_Q\left[ \zeta, X,T\right] e^{i\int\;dt\;\zeta\left( t\right)u\left( t\right)/\hbar }
\te
where 

\be
P_Q\left[ \zeta, X,T\right] =e^{i\mathcal{L}\left(T\right)}P\left[ \zeta, X\right] 
\te

\be
\mathcal{L}\left(T\right)=\frac{m^2}{4\hbar}\int\;dt\;\left(\Omega^2\right)'X\left(t\right)\left[i\hbar\frac{\delta}{\delta\zeta\left( t\right) } \right] 
\int\;dt'\;\left(\Omega^2\right)'\left(t'\right)D\left(t,t'\right)\left[i\hbar\frac{\delta}{\delta\zeta\left( t'\right) } \right]^2
\te
We now put all together

\bea 
&&\rho_{Sf}\left( X_f+\frac {u_f}2,X_f-\frac {u_f}2,T\right) =\int_{X\left( t_i\right) =X_i,u\left( t_i\right) =u_i}^{X\left( T\right) =X_f,u\left( T\right) =u_f} DXDu\;\int D\zeta\;\mathcal P_Q\left[ \zeta, X,T\right] e^{\left( i/\hbar\right)u_f\dot X_f}\nn
&&e^{\left( -im/\hbar\right) \int dt\;u\left( t\right)\left\lbrace m\ddot X\left(t\right)+ \mathcal{D}\left[ X\right] -\zeta\right\rbrace \left( t\right)}e^{\left( -im/\hbar\right)u_i\dot X_i}  \rho_{Si}\left( X_i+\frac {u_i}2,X_i-\frac {u_f}i,t_i\right)
\tea 
where 

\be
\mathcal{D}\left[ X\right]\left(t\right)=\mathcal{D}_0\left[ X\right]\left(t\right)+\mathcal{D}_2\left[ X\right]\left(t\right)
\te 

\bea
\mathcal{D}_0\left[ X\right]\left(t\right)&=& m\Omega^2\left[ \Theta\right]X\left(t\right)\nn
\mathcal{D}_2\left[ X\right]\left(t\right)&=&-m^2\left(\Omega^2\right)'X\left(t\right)
\int\;dt'\;\left(\Omega^2\right)'\left(t'\right)D\left(t,t'\right)X^2\left( t'\right)
\tea

\section{The system's Wigner function}
Although the reduced density matrix gives a full description of the quantum state of the system, its dynamics, given by the so-called master equation, is rather involved \cite{Book}. It is more heuristic to introduce the Wigner function \cite{Hillery,Cosmas}

\be 
F_W\left( X,P,t\right) =\int du\;e^{-iPu}\rho_S\left( X+\frac u2,X-\frac u2,t\right) 
\te 
Introducing

\be 
1=\int DP\;\delta\left( P-m\dot X\right) 
\te 
into the path integral, we get 

\bea 
&&F_W\left( X_f,P_f,T\right) =\int_{X\left( t_i\right) =X_i,P\left( t_i\right) =P_i}^{X\left( T\right) =X_f,P\left( T\right) =P_f} DXDP\;\int D\zeta\;\mathcal P_Q\left[ \zeta, X,T\right] \nn
&&\delta\left( \dot X-\frac Pm\right) \delta \left( \dot P+  \mathcal{D}\left[ X\right] -\zeta\right)f\left( X_i,P_i,t_i\right) 
\tea 
The Wigner function obeys the equation (see \cite{Book} and Appendix A)

\bea 
\frac{\partial F_W}{\partial t}+\left\lbrace H_{AS},F_W\right\rbrace &=&\frac{\partial}{\partial P}\left\lbrace \mathcal{D}_2F_W-\int dt'\bar{\mathcal{N}}\left( t,t'\right)  \left\lbrace \bar X\left( t'\right) ,F_W\right\rbrace \right.\nn
&+&\left.\frac{m^2\hbar^2}{4}\left(\Omega^2\right)'X
\int dt' D\left(t,t'\right)\left(\Omega^2\right)'\left( t'\right) \left\lbrace\bar X\left( t'\right),\left\lbrace \bar X\left( t'\right),F_W\right\rbrace \right\rbrace \right\rbrace 
\label{Wignereq}
\tea
where $\bar X\left(t'\right)$ is the solution to the equations

\bea 
 \frac d{dt} \bar X&=&\frac {\bar P}m\nn
 \frac d{dt} \bar P&=&-  \mathcal{D}\left[\bar X\right]+\zeta
\tea 
which passes through $\left(X,P\right)$ at time $t$. $\zeta$ is multiplicative noise with distribution function $P_Q\left[ \zeta, \bar X,t\right]$.
 
The idea is to solve this equation perturbatively, first the homogeneous solution, then replacing $F_W$ by the homogeneous solution in the right hand side, and so on.

It helps to notice that one can make a canonical transformation (see Appendix B)

\bea
X&=&\sqrt{\frac J{m}}\left[ f\left( t\right) e^{-i\phi}+f^*\left( t\right) e^{i\phi}\right] \nn
P&=&\sqrt{{mJ}}\left[ \dot f\left( t\right) e^{-i\phi}+\dot f^*\left( t\right) e^{i\phi}\right] 
\label{canon}
\tea 
where $f$ and $f^*$ are WKB wavefunctions.  $J$ is the adiabatic invariant of the linearized dynamics. Because this is an exact solution of the linear dynamics when $J$ and $\phi$ are constant, the homogeneous solution is just any time-independent function of $J$ and $\phi$. The initial energy is $E_0=\omega\left( 0\right) J$; if the initial state is the $n$-th energy eigenstate then the zeroth order solution $F_{Wn}$ depends only on $J$. The first order term reads

\bea
&&\delta F_{W}=\frac{m^2}2\int dt \left(\Omega^2\right)'\left( t\right)\int\;dt'\;\left(\Omega^2\right)'\left(t'\right)\left[D\left(t,t'\right)\left\{X^2,
X^2\left( t'\right)F_{Wn}\right\}\right.\nn
&+&\frac12N\left( t,t'\right) \left\{X^2,  \left\lbrace  X^2\left( t'\right) ,F_{Wn}\right\rbrace\right\} \nn
&-&\left.\frac{\hbar^2}{4} D\left(t,t'\right)\left\{X^2,\left\lbrace X\left( t'\right),\left\lbrace  X\left( t'\right),F_{Wn}\right\rbrace \right\rbrace \right\rbrace\right] 
\tea
$\delta F_{W}$ describes a non-stationary state

\be
\delta F_{W}=\sum_{k=-4}^4F_{W}^{\left(k\right)}\left(J,t\right)e^{ik\phi}
\te
However, if we measure the energy at the end of the protocol the state collapses onto its $\phi$- independent part

\be 
F_{W}^{\left(0\right)}=\nu J\left[F'_{Wn}+\left(JF'_{Wn}\right)'\right]+\mu \frac J{\hbar}\left[F_{Wn}+\left(JF_{Wn}\right)'+\frac{\hbar^2}{4}\left[F''_{Wn}+\left(JF''_{Wn}\right)'\right]\right]
\label{finalstate}
\te
where

\bea
\nu&=&\int dt \left(\Omega^2\right)'\left( t\right)\int\;dt'\;\left(\Omega^2\right)'\left(t'\right)N\left( t,t'\right) \left[f^2\left( t\right)f^{*2}\left( t'\right)+\mathrm {cc}\right]\nn
&=&2\left\langle \vartheta\left(0\right)^2\right\rangle\left|I_0\right|^2+2M^{-2}\left\langle P\left( 0\right)^2\right\rangle \left|I_1\right|^2\nn
\mu&=&\frac{i\hbar}4\int dt \left(\Omega^2\right)'\left( t\right)\int\;dt'\;\left(\Omega^2\right)'\left(t'\right)D\left(t,t'\right)
\left[f^2\left( t\right)f^{*2}\left( t'\right)-\mathrm {cc}\right]\nn
&=&\frac {i\hbar}{8M}\left[I_1I_0^*-I_0I_1^*\right]
\label{coefs}
\tea

\bea
I_0&=&\int dt \left(\Omega^2\right)'\left( t\right)f^2\left( t\right)\nn
I_1&=&\int dt\;t \left(\Omega^2\right)'\left( t\right)f^2\left( t\right)
\label{jintegrals}
\tea
To compute the final state, we use the recurrence relations for the Wigner functions of a harmonic oscillator (see \cite{Cosmas} and Appendix C)

\bea
JF_{Wn}&=&\frac{\hbar}4\left[\left(2n+1\right)F_{Wn}+nF_{W\left(n-1\right)}+\left(n+1\right)F_{W\left(n+1\right)}\right]\nn
JF'_{Wn}&=&\frac12\left[nF_{W\left(n-1\right)}-F_{Wn}-\left(n+1\right)F_{W\left(n+1\right)}\right]\nn
\frac{\hbar J}4F''_{Wn}&=&-\left(n+\frac12\right)F_{Wn}+\frac J{\hbar}F_{Wn}-\frac{\hbar}4F'_{Wn}
\label{rrs}
\tea
We obtain

\be 
F_{W}^{\left(0\right)}=\frac12\left[\left(\nu +\frac{\mu}2\right)n\left(n-1\right)F_{W\left(n-2\right)}+\left(\nu -\frac{\mu}2\right)\left(n+1\right)\left(n+2\right)F_{W\left(n+2\right)}-2\left(\nu\left(1+n+n^2\right)-\mu \left(2n+1\right)\right)F_{W}^{\left(0\right)}\right]
\te
We see that within this approximation, after measuring the energy in the final state, the system may have undergone a transition to the $n\pm 2$ states, thus violating adiabaticity.

In particular, the variation in the mean occupation number is

\bea
\Delta n&=&\left(\nu -\frac{\mu}2\right)\left(n+1\right)\left(n+2\right)-\left(\nu +\frac{\mu}2\right)n\left(n-1\right)\nn
&=&2\nu\left(2n+1\right)-{\mu}\left(n^2+n+1\right)
\tea
The extra energy injected into the system is $\delta W=\Delta n\hbar\omega\left( T\right)$. The condition for adiabaticity is $\Delta n\ll 1$

\section{Estimating the cost of the STA} 
To conclude our analysis we must estimate the coefficients $\nu$ and $\mu$ from eq. (\ref{finalstate}).

Observe that $\mu$ depends on the inertia of the driving but not on its quantum state, while $\nu$ depends on both.

\begin{figure}
\includegraphics[scale=1]{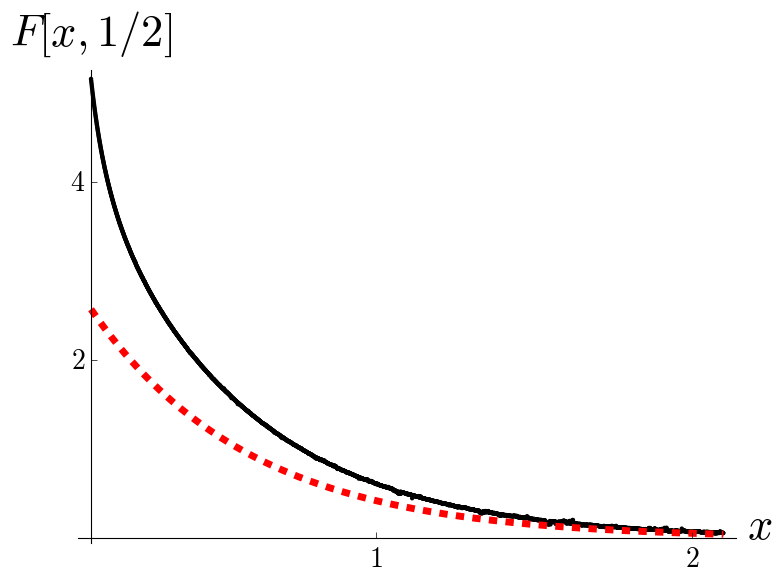}
\caption{[Color online] (full line) numerical evaluation of $F\left[x,1/2\right]$ from eq. (\ref{full}); (dashes) the asymptotic form for $\delta\to 0$, eq. (\ref{asym})}
\label{fxy}
\end{figure}

Actually, if $d\Theta/dt=$ constant throughout the protocol, then $I_0=0$ (see Appendix D), so $\mu=0$, and 

\be
\nu= 2M^{-2}\left\langle P_D\left( 0\right)^2\right\rangle \left|I_1\right|^2
\te
where, from eq. (\ref{jintegral1})

\be
I_1=\frac1{\dot\Theta}\int dt\; \left[-i\left(\frac{\dot\omega}{\omega}\right)-\frac1{4}\left(\frac{\dot\omega^2}{\omega^3}\right)\right]\left( t\right)e^{-2i\int\omega dt}
\label{jintegral2}
\te
To see that this is generally nonzero, consider a protocol of the form

\be
\omega=\omega_0+\delta\arctan\left(t/\tau\right)
\label{protocol}
\te
$\delta/\omega_0\le 2/\pi$. The condition that $\Omega^2\ge 0$ requires $\omega_0\tau\ge\sqrt{3/4}\left(\delta/\omega_0\right)$
Then

\be
\dot\omega=\frac{\delta}{\tau}\frac1{1+\left(\frac t{\tau}\right)^2}
\te
while

\be
\int\omega dt=\omega_0\tau\left\lbrace \left(1+\frac{\delta}{\omega_0}\arctan\left(\frac t{\tau}\right)\right)\frac t{\tau}-\frac{\delta}{2\omega_0}\ln\left[1+\left(\frac t{\tau}\right)^2\right]\right\rbrace 
\te
So, separating the dimensionful constants, we find 

\be
\left|I_1\right|=\frac{\delta}{\dot\Theta\omega_0}F\left[\omega_0\tau,\frac{\delta}{\omega_0}\right]
\te

\be 
F\left[ x,y\right] =\left|\int\frac{ ds\left[ 1-\frac{i\left( y/4x\right) }{\left( 1+s^2\right) \left(1+y\arctan\left(s\right)\right)^2}\right]}{\left( 1+s^2\right) \left(1+y\arctan\left(s\right)\right)} e^{-2ix\left\lbrace \left(1+y\arctan\left(s\right)\right)s-\frac{y}{2}\ln\left[1+s^2\right]\right\rbrace }\right|
\label{full}
\te
In figure (\ref{fxy}) we plot $F\left[x,1/2\right]$, together with the asymptotic form (compare with \cite{Landau32a,Landau32b})

\be
F\left[x,0\right]=\pi\;e^{-2x}
\label{asym}
\te
in the range $x\ge 0.1$. For smaller $x$ at fixed $y\not=0$ we find $F\propto 1/x$. Finally

\be
\nu= 2F^2\left[\omega_0\tau,\frac{\delta}{\omega_0}\right]\left(\frac{\delta}{\omega_0}\right)^2\frac{\left\langle P_D\left( 0\right)^2\right\rangle}{\left(M\dot\Theta\right)^2}
\te
Since we expect the driving to complete the protocol in a time $\approx\tau$, $\left\langle P_D^2\right\rangle/2M\approx\delta E$ is constrained by the quantum speed limit \cite{Tamm,Funo17,Lloyd,Deffner17}. Adopting a simple uncertainty relation estimate $\left\langle P_D^2\right\rangle/2M\approx\delta E\ge \hbar/\tau$, and also estimating $M\dot\Theta^2/2\approx H_D\left[ \bar\Xi\right] $ as the energy of the driving system (cfr. eq. (\ref{universe})), we get

\be
\nu\approx 2{F^2\left[\omega_0\tau,\frac{\delta}{\omega_0}\right]}\left(\frac{\delta}{\omega_0}\right)^2\frac{\hbar}{H_D\left[ \bar\Xi\right]\tau}
\te
Therefore, to make $\nu$ small we must make $\tau$ large.

\section{Final remarks}
It is known that shortcuts to adiabaticity have definite costs in terms of work that has to be done during the protocol, although within the classical approximation the work invested is recovered at the end  \cite{Zheng16,Campbell17}. We have shown that there is also a definite cost in terms of particles being created during the protocol, so that there is a violation of adiabaticity. To lowest nontrivial order the mean number of particles created is $\Delta n=2\nu\left(2n+1\right)$, where

\be 
\nu\approx 2\pi^2\left(\frac{\delta}{\omega_0}\right)^2\frac{\hbar}{H_D\left[ \bar\Xi\right]\tau}e^{-4\omega_0\tau}
\te
The meaning of $\omega_0$, $\delta$ and $\tau$ comes from the explicit protocol eq. (\ref{protocol}), and $H_D\left[ \bar\Xi\right]$ is the energy for the classical trajectory of the driving system, see eq. (\ref{universe}). For a given driving system energy, $\nu$ can be made arbitrarily small only in the adiabatic limit.

This deviation from adiabaticity is intrinsic in the sense it depends on no parameter not present under the classical approximation. It has no analog in a classical model, where to obtain a similar result one has to assume the driving is at finite temperature, it has a definite dissipation mechanism, or both. Of course in practice these classical sources of noise and dissipation are likely to overwhelm the effect we have discussed, but we believe the fact that such an intrinsic deviation from adiabaticity exists is relevant from a first principles point of view.

Actually, the picture emerging from our analysis is quite simple. Quantum fluctuations in the driving cause uncertainty in the initial value of the driving's coordinate and its initial speed. If we had simply replaced these uncertain quantities by $c-$ number random variables with the proper distribution, we would have arrived essentially to the same final result in a much more direct way. However, it is unclear such a replacement is justified, because quantum fluctuations mediate interactions and provide dissipative mechanisms, beyond rattling the system. So we went through a systematic derivation of the lowest nontrivial order result, keeping all relevant terms.

Of course, to have a detailed derivation such as this may be useful to go beyong this leading order result or to seek similar effects in other types of quantum engines. We expect to deal with these matters in future communications.

Meanwhile, once the deviation from adiabaticity is properly characterized, it may be possible to compensate for it. In fact, some relevant steps seem to have been taken already \cite{Muga18,CKHu18}.

\acknowledgements 
Work supported in part by Universidad de Buenos Aires and CONICET (Argentina).

It is a pleasure to acknowledge exchanges with A. del Campo, N. Freitas, M. Larocca, J. G. Muga and D. Wisniacki.

\section*{Appendix A: derivation of eq. (\ref{Wignereq})}

To simplify the analysis to follow, it is convenient to discretize time. Write $X_k=X\left(t_k\right) $, $t_k=t_i+kdt$, and so on. Then

\be 
F_{W \left(k+1\right)}\left( X,P\right) =\left\langle \delta\left( X-X_{k+1} \right)  \delta\left( P-P_{k+1} \right)\right\rangle 
\te 
where

\bea 
 X_{j+1}&=&X_j+dt\frac {P_j}m\nn
 P_{j+1}&=&P_j+dt\left[-\mathcal{D}_j+\zeta_j\right]
\tea 

\be
\mathcal{D}_j= m\Omega^2\left[ \Theta_j\right]X_j-m^2dt\left(\Omega^2\left[\Theta_j\right]\right)'X_j\sum_{l=0}^{j-1}D\left(t_j,t_l\right)
\left(\Omega^2\left[\Theta_l\right]\right)'X_l^2
\te
The average is over initial conditions, weighted by $F_{W0}\left(X_0,P_0\right)$, and over the noise with distribution

\be
P_{Q\left(j+1\right)}=e^{i \mathcal{L}_{j+1}}P_{j+1} 
\te

\be
\mathcal{L}_{j+1}=\mathcal{L}_{j}+\mathcal{L}_{j,j+1}
\te

\be
\mathcal{L}_{j,j+1}=
\frac{m^2}{4\hbar dt}\left(\Omega^2\left[\Theta_j\right]\right)'X_j\left[i\hbar\frac{\partial}{\partial\zeta_j } \right] 
\sum_{m=0}^{j-1}D\left(t_j,t_m\right)\left(\Omega^2\left[\Theta_m\right]\right)'\left[i\hbar\frac{\partial}{\partial\zeta_m } \right]^2
\te

\be
\mathcal{N}_{kl}=m^2\left(\Omega^2\left[\Theta_k\right]\right)'X_k N\left( t_k,t_l\right) \left(\Omega^2\left[\Theta_l\right]\right)' X_l
\te

\be
P_{j+1}=P_j\frac{e^{-\left(\zeta_j-\sum_{p=0}^{j-1}\gamma_{jp}\zeta_p\right)^2/2dt^2\gamma_j}}{\sqrt{2\pi dt\gamma_j^{1/2}}}
\te
where

\be
\sum_{m=0}^{j-1}\gamma_{jm}\mathcal{N}_{mq}=\mathcal{N}_{jq}
\te

\be
\gamma_j=\mathcal{N}_{jj}-\sum_{m=0}^{j-1}\sum_{r=0}^{j-1}\gamma_{jm}\mathcal{N}_{mj}
\te
In particular

\be
\int d\zeta_j\;P_{j+1}=P_j
\te
We now write the Wigner function as

\be
F_{W\left(j+1\right)}\left(X,P\right)=\int dX_0dP_0\;F_{W0}\left(X_0,P_0\right)\int\prod_{k=0}^jd\zeta_k\;P_{Q\left(j+1\right)}\delta\left(X-X_{j+1}\right)\delta\left(P-P_{j+1}\right)
\te
Because of the causal prescription we have chosen, we avoid the appearance of a Jacobian within the path integral.

The idea is to elliminate $\zeta_j$. From the recursion relation

\bea 
 X_{j+1}&=&X_j+dt\frac {P_j}m\nn
 P_{j+1}&=&P_j+dt\left[-\mathcal{D}_j+\zeta_j\right]
\tea 
to first order we get

\bea
\delta\left(X-X_{j+1}\right)\delta\left(P-P_{j+1}\right)&=&\delta\left(X-X_j-dt\frac {P_j}m\right)\delta\left(P-P_j+dt\left[\mathcal{D}_j-\zeta_j\right]\right)\nn
&=&\left[1-dt\frac{\partial}{\partial X}\frac {P}m+dt\frac{\partial}{\partial P}\left[\mathcal{D}_j-\zeta_j\right]\right]\delta\left(X-X_{j}\right)\delta\left(P-P_{j}\right)
\tea
Whereby the $\zeta_j$ dependence is made explicit. Next we integrate by parts

\bea
F_{W\left(j+1\right)}\left(X,P\right)&=&\int dX_0dP_0\;F_{W0}\left(X_0,P_0\right)\int\prod_{k=0}^jd\zeta_k\;e^{i \mathcal{L}_{j}+\mathcal{L}_{j,j+1}}P_{j+1} \delta\left(X-X_{j+1}\right)\delta\left(P-P_{j+1}\right)\nn
&=&\int dX_0dP_0\;F_{W0}\left(X_0,P_0\right)\int\prod_{k=0}^jd\zeta_k\;e^{i \mathcal{L}_{j}}P_{j+1} e^{-i\mathcal{L}_{j,j+1}}\delta\left(X-X_{j+1}\right)\delta\left(P-P_{j+1}\right)\nn
&=&F_{Wj}\left(X,P\right)+dt\left[A+B+C+D\right]
\tea
where

\bea
A&=&-\frac{\partial}{\partial X}\frac {P}mF_{Wj}\left(X,P\right)\nn
B&=&\frac{\partial}{\partial P}\int dX_0dP_0\;F_{W0}\left(X_0,P_0\right)\int\prod_{k=0}^jd\zeta_k\;P_{Q\left(j\right)}\mathcal{D}_j\delta\left(X-X_{j}\right)\delta\left(P-P_{j}\right)\nn
C&=&-\frac{\partial}{\partial P}\int dX_0dP_0\;F_{W0}\left(X_0,P_0\right)\int\prod_{k=0}^jd\zeta_k\;P_{Q\left(j\right)}\left[\sum_{p=0}^{j-1}\gamma_{jp}\zeta_p\right]\delta\left(X-X_{j}\right)\delta\left(P-P_{j}\right)\nn
D&=&\frac{\partial}{\partial P}\int dX_0dP_0\;F_{W0}\left(X_0,P_0\right)\int\prod_{k=0}^jd\zeta_k\;P_{Q\left(j\right)}\left[i\mathcal{L}_{j,j+1}\zeta_j\right]\delta\left(X-X_{j}\right)\delta\left(P-P_{j}\right)\nn
\tea
The idea is to compute these terms to lowest nontrivial order in $\Omega$ derivatives. The $A$ term is already in its final form. In the $B$ term we replace $\mathcal{D}$ by $\bar{\mathcal{D}}$, namely we evaluate it on the linearized trajectory $\left( \bar X_k,\bar P_k\right) $. In $C$ and $D$ we neglect $\mathcal{L}_j$ and we approximate $P_j$ by $\bar{P}_j$. Then

\bea 
C&=&-\frac{\partial}{\partial P}\int dX_0dP_0\;F_{W0}\left(X_0,P_0\right)\int\prod_{k=0}^jd\zeta_k\;P_{Q\left(j\right)}\left[\sum_{p=0}^{j-1}\gamma_{jp}\zeta_p\right]\delta\left(X-X_{j}\right)\delta\left(P-P_{j}\right)\nn
&=&\frac{\partial}{\partial P}\left[\frac{\partial}{\partial X}\sum_{p=0}^{j-1}\bar{\mathcal{N}}_{jp} \frac{\partial X_j}{\partial \zeta_p}+\frac{\partial}{\partial P}\sum_{p=0}^{j-1}\bar{\mathcal{N}}_{jp} \frac{\partial P_j}{\partial \zeta_p}\right] F_{Wj}\left(X,P\right)
\tea
The derivatives are computed as in \cite{Book}

\bea 
 \frac{\partial X_j}{\partial \zeta_p}&=&-dt \frac{\partial \bar X_{p+1}}{\partial P_j}\nn
 \frac{\partial P_j}{\partial \zeta_p}&=&\;dt \frac{\partial \bar X_{p+1}}{\partial X_j}
\tea
Introducing the Poisson brackets

\be 
\left\lbrace f,g\right\rbrace = \frac{\partial f}{\partial P}\frac{\partial g}{\partial X}- \frac{\partial f}{\partial X}\frac{\partial g}{\partial P}
\te 

\be 
C=-\frac{\partial}{\partial P}dt\sum_{p=0}^{j-1}\bar{\mathcal{N}}_{jp} \left\lbrace \bar X_{p+1},F_{Wj}\left(X,P\right)\right\rbrace 
\te 
In the same way

\be 
D=\frac{m^2\hbar^2}{4}\frac{\partial}{\partial P}\left(\Omega^2\left[\Theta_j\right]\right)'X_j
dt\sum_{p=0}^{j-1} D\left(t_j,t_p\right)\left(\Omega^2\left[\Theta_p\right]\right)'\left\lbrace\bar X_{p+1},\left\lbrace \bar X_{p+1},F_{Wj}\right\rbrace \right\rbrace 
\te
We may now take the continuum limit, whereby we find eq. (\ref{Wignereq}).

\section*{Appendix B: Eq. (\ref{canon}) as a canonical transformation}
It is rather essential to show that the transformation eq. (\ref{canon}) is canonical, since only then we can compute Poisson brackets indistinctly in either set of variables, $\left( P,X\right) $ or $\left( J,\phi\right) $. 

We need to show that there is a function $G=G\left( X,\phi\right) $ such that

\be 
PdX-H_{AS}dt=Jd\phi-Kdt-dG
\te 
where $K$ is the Hamiltonian in the new variables; this means

\bea 
\frac{\partial G}{\partial X}&=&-P\nn
\frac{\partial G}{\partial \phi}&=&J\nn
\frac{\partial G}{\partial t}&=&H_{AS}-K
\tea 
In our case 

\be 
G=-\frac12mX^2\left\lbrace \frac{ \dot f\left( t\right) e^{-i\phi}+\dot f^*\left( t\right) e^{i\phi}}{f\left( t\right) e^{-i\phi}+f^*\left( t\right) e^{i\phi}} \right\rbrace 
\te 
From the osillator equation for $f$ we see that actually $K=0$, so both $J$ and $\phi$ are constants of motion.

\section*{Appendix C: Harmonic oscillator Wigner functions}
For a harmonic oscillator in an energy eigenstate, the Wigner function obeys

\be
\left[\frac{-\hbar^2}{2m}\frac{\partial^2}{\partial x^2}+\frac12m\omega^2x^2-\hbar\omega\left(n+\frac12\right)\right]\int\frac{dp}{h}e^{ip\left(x-y\right)}F_{Wn}\left(X,p\right)=0
\te
where $X=\left(x+y\right)/2$. Observe that

\be
x^2=\left(X+\frac{x-y}2\right)^2=X^2+X\left(x-y\right)+\frac14\left(x-y\right)^2
\te
then we get

\be
\left\{\frac{-1}{2m}\left[-p^2+i\hbar p\frac{\partial}{\partial X}+\frac{\hbar^2}4\frac{\partial^2}{\partial X^2}\right]+\frac12m\omega^2\left[X^2+i\hbar X\frac{\partial}{\partial p}-\frac{\hbar^2}4\frac{\partial^2}{\partial p^2}\right]-\hbar\omega\left(n+\frac12\right)\right\}F_{Wn}=0
\te
or else

\be
\left[\omega J-\hbar\omega\left(n+\frac12\right)\right]F_{Wn}-\frac i2\hbar\omega\left\{J,F_{Wn}\right\}-\frac{\hbar^2}4\left[\frac1{2m}\left\{p,\left\{p,F_{Wn}\right\}\right\}+\frac12m\omega^2\left\{X,\left\{X,F_{Wn}\right\}\right\}\right]=0
\te
Recall that

\bea
X&=&\sqrt{\frac {2J}{m\omega}}\cos\left(\phi+\omega t\right) \nn
p&=&-\sqrt{2m\omega J}\sin\left(\phi+\omega t\right) 
\tea
so, $F_{Wn}=F_{Wn}\left(J\right)$ and

\be
\frac{\hbar}{4}\left[JF_{Wn}''+F_{Wn}'\right]+\left[n+\frac12-\frac J{\hbar}\right]F_{Wn}=0
\label{eq}
\te
Write

\be
s=\frac{4J}{\hbar}
\te
and 

\be
F_{Wn}=e^{-s/2}L_n\left[s\right]
\te
Then

\be
s\ddot L_n+\left[1-s\right]\dot L_n+nL_n=0
\te
$L_n$ is a Laguerre polynomial \cite{Lebedev}. 

The normalization is determined by

\be
f_n\left(0\right)=\int dy\;\psi_n\left(y/2\right)\psi_n^*\left(-y/2\right)=2\left(-1\right)^n
\te
Since $L_n\left(0\right)=1$ to get this we must define

\be
f_n=2\left(-1\right)^ne^{-s/2}L_n\left[s\right]
\te
Eqs.(\ref{rrs}) follow from eq. (\ref{eq}) and the recursion relations for Laguerre polynomials

\bea
\dot L_n-\frac ns\left[L_n-L_{n-1}\right]&=&0\nn
\left(n+1\right)L_{n+1}+\left[s-2n-1\right]L_{n}+nL_{n-1}&=&0
\tea

\section*{Appendix D: $I_0$ and $I_1$ when $d\Theta/dt=$ constant}
We want to show that when $d\Theta/dt=$ constant, $I_0=0$ and $I_1$ may be greatly simplified. The point is that up to a constant, we may replace the $\Theta$ derivative by a time derivative in eqs. (\ref{jintegrals}). From eq. (\ref{WKB2})

\bea
\frac d{dt}\Omega^2&=&2\omega\left[\dot\omega+\frac1{4\omega}\frac d{dt}\left(\frac{\ddot\omega}{\omega}-\frac32\left(\frac{\dot\omega}{\omega}\right)^2\right)\right]\nn
&=&2\omega\left[\dot\omega+\frac1{4}\frac d{dt}\frac1{\omega}\frac d{dt}\left(\frac{\dot\omega}{\omega}\right)\right]
\label{WKB3}
\tea 
$I_0=0$ follows from a double integration by parts of the second term. With respect to $I_1$, we find

\bea
I_1&=&\frac1{\dot\Theta}\int dt\;t \left[\dot\omega+\frac1{4}\frac d{dt}\frac1{\omega}\frac d{dt}\left(\frac{\dot\omega}{\omega}\right)\right]\left( t\right)e^{-2i\int\omega dt}\nn
&=&\frac1{\dot\Theta}\int dt\; \left[-i\left(\frac{\dot\omega}{\omega}\right)-\frac1{4}\left(\frac{\dot\omega^2}{\omega^3}\right)\right]\left( t\right)e^{-2i\int\omega dt}
\label{jintegral1}
\tea
namely eq. (\ref{jintegral2})

\end{document}